\documentclass[vecphys]{svmult}

\usepackage{makeidx}         
\usepackage{graphicx}        
\usepackage{multicol}        
\usepackage{natbib}          

\makeindex             

\newcommand{\vsh}{\mbox{$V_{\rm shock}$}}
\newcommand{\kms}{km~s$^{-1}$}
\newcommand{\hm}{H$_2$}

\begin{document}

\title*{Hypersonic Molecular Shocks in Star Forming Regions}
\author{P. W. J. L. Brand\inst{1}}
\institute{Institute for Astronomy, University of Edinburgh
\texttt{pwb@roe.ac.uk}}
%
%
\maketitle


\section{A Little History}
\label{sec:history}

Shocks are everywhere in the Universe, and with hindsight it is
easy to see why.

In the John Dyson era one might be tempted to ask if indeed there
is anything else? Such a coherent view of so much of the goings-on
between the stars has been imparted by John and his colleagues
that most interesting phenomena now have a Dyson -- or Dyson-style
-- similarity solution to describe them.

But the discovery of shocks in the cold clouds where stars are
born caused widespread surprise.

The earliest clear indications came in a rush. As noted in the chapter
by Tom Ray, it was pointed out by \citet{Sc75} that the enigmatic
Herbig-Haro (HH) objects were shock-excited; and in 1976, nearly
simultaneously, the new techniques of millimetre and infrared
astronomy led to important discoveries: 2.3mm CO line profiles with a
width of 150 \kms\ were observed by \citet{Kw76} and by \citet{Zu76}
around the position of the Orion IR cluster, and in a similar location
infrared spectra obtained by \citet{Ga76} showed \hm\ quadrupole
lines, revealing molecular gas with an apparent temperature of
$2000\;$K~\citep{Be78}.

Very soon a detailed map of shock excited \hm\ by \citet{Be79}
appeared, showing two lobes centred around the Becklin
Neugebauer/Kleinmann Low infra-red cluster; and the important
measurement by \citet{Na79} of extremely large \hm\ line-width in the
Orion outflow set the scene.

Almost immediately several important models were
published~\citep{HS77,Lo77,Kw77}. \citet{Kw77} demonstrated that a
hydrodynamic shock travelling faster than 24 \kms\ through molecular
gas would completely dissociate the \hm. This created a major
difficulty in explaining the wide observed profiles.

The two-fluid magnetic shock (`C type') model introduced by
\citet{Dr80} and \citet{DRD83} was far less destructive to \hm\  and
provided reasonable fits to the line intensities measured at that
time~\citep{Ch82,DR82}.

\begin{figure}
\centering
\includegraphics[height=8cm]{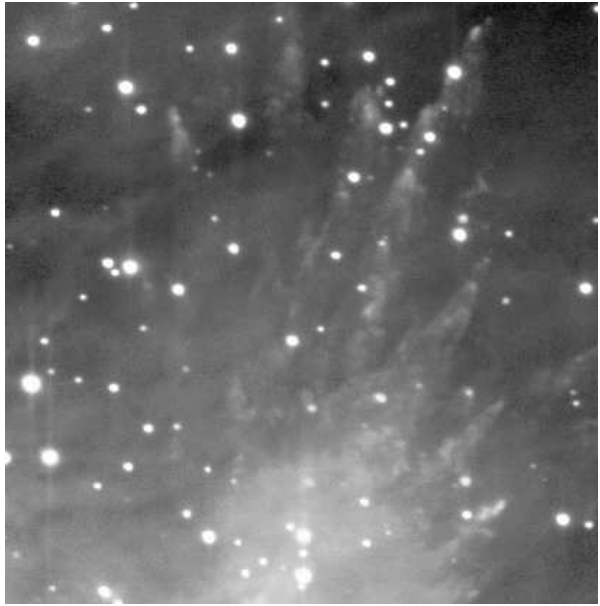}
\caption{The northern part of the Orion molecular outflow  showing
the \hm\ emission in a series of bow shocks. The bright tips show
up in the light of [FeII], due to strong J shocks. [IRIS at AAT
picture, courtesy M.G. Burton] } \label{fig:OrionShock}
\end{figure}

Meanwhile the hunt for hot \hm\  raced on. Many HH objects were
observed: \citet{Mu87} listed fourteen such sources;
and the Orion outflow, the brightest source, had been studied in
detail.

A plot of column densities from the near IR lines of \hm\ measured
in the Orion outflow \citep{Br88} demonstrated that a single C type
shock could not explain the excitation there, and maps of line
ratios \citep{Br89} in the outflow, showing little change in
excitation with position, further clouded the issue.

As well as these star-forming regions, supernova remnants were
mapped in shock-excited \hm. The old remnant IC443 shows shocked
\hm\  emission round a large part of the shell, and the line
excitation appears strikingly similar to that found in
Orion~\citep{Gr87,Mo91}. More surprisingly, \hm\  was found in the
heart of the Cygnus loop~\citep{Gr91}.

In his review \citet{Mu87} suggested one possible
explanation for the \hm\  emission often found near, but not
exactly at the optical knots in HH objects (there was initially
some uncertainty because of the low spatial resolution of the
early IR measurements). Amongst other possibilities he proposed
that the \hm\ excitation might come from the flanks of a bow
shock, the head of which produced the optical emission.

In this he was following on from the original
\citet{Sch78} suggestion that a bow shock model might be
used to explain the range of different excitations seen in the
optical spectra of HH objects. Subsequently, \citet{HR84}
proposed that a bow shock model could replicate the observations
of HH1 and HH2; several papers demonstrated that bow
shock model position/velocity diagrams fitted the
observations~\citep{Ch85,RB85,Ra86}; and the important
paper by \citet{Ha87} explored in
detail the excitation and line profile predicted for bow shock
observations.

This idea was applied to the observations of the Orion
outflow~\citep{Sm90a, Sm90b,Sm91b}, demonstrating that single plane
C-shocks could not explain the observations, but that a C shock
propagating into material with a very high magnetic field could do
so. The magnetic field required was extremely high, implying an
unusual pre-shock magnetic field.

The review by \citet{DMcK93} summarised the
theoretical and observational state at the time.

Before going further, we shall peruse the properties of these
shocks.

\section{Shock Basics}
\label{sec:basics}

\subsection{Hydrodynamic Jumps, Dissociation, Isothermal Shocks, Magnetic Fields}
\label{subsec:jumps}

As a prelude, we examine the simplest possible case of a steady
strong plane shock in an ideal gas~\citep[e.g.,][]{DMcK93}.

Consider the frame in which the shock front is static and the
upstream gas flows perpendicularly towards it, at a speed \vsh\
much greater than the random thermal motions -- hence
hypersonically. This is a parallel flow of independent particles
which encounters a wall of more slowly moving post-shock material.
Each incoming particle collides elastically (within one or two
mean free paths) with a post-shock particle, sharing its energy
and randomizing its direction of travel. Thus, downstream of the
shock the gas is hot ($\frac{3}{2}kT \simeq \frac{1}{2}m\vsh^2$) and, because
of the redirection of velocity, slower.

It is evident that the shock is a `sudden' transition in the
hydrodynamic sense, since hydrodynamics is applicable only to
scales much greater than a mean free path for elastic collisions;
and it remains sudden because to smooth itself it would have to
propagate pressure waves upstream at the speed of sound through
gas arriving at a speed much greater than that.

At a downstream distance considerably larger than the elastic mean
free path (in fact of the order of the {\em in}elastic mean free
path) there has been time for collisions to excite the internal
states of the gas (which could for example lead to ionization
and/or dissociation if the shock is strong enough) and to change
the effective $\gamma$ from 5/3 to a smaller number $N+5/N+3$
where $N$ is the effective number of internal degrees of freedom
of the gas particles. The internal states then de-excite by
radiating photons and the gas cools. The size of this cooling zone
depends on the rate of this process, and is large compared with
the mean free path but usually small compared with the global
structure of the shocked region.

More formally, in this frame the mass, momentum and energy are
conserved through the front. We denote upstream gas by suffix `0',
downstream by `1'; $\rho$ is density, $p$ is pressure and $v$ is
velocity in this frame, and we assume that the gas is ideal with
constant adiabatic index $\gamma=5/3$. This value corresponds to
velocity thermalization by elastic collisions but no internal
excitation. Hence, the Rankine-Hugoniot shock jump conditions are:
\begin{eqnarray*}
\rho_1v_1 & = & \rho_0\vsh \\
p_1 + \rho_1v_1^2 & = & p_0 + \rho_0\vsh^2 \\
\frac{\gamma}{\gamma-1}\frac{p_1}{\rho_1} + \frac{1}{2}v_1^2 & = &
\frac{\gamma}{\gamma-1}\frac{p_0}{\rho_0} + \frac{1}{2}\vsh^2
\end{eqnarray*}

Further behind the shock when cooling sets in at a net rate
$\mathcal{L}$ W kg$^{-1}$, the first two equations, expressing
conservation of mass and momentum, continue to apply while the
third becomes the initial condition for
\[ v\frac{d}{dx}\left(\frac{\gamma}{\gamma-1}\frac{p_1}{\rho_1} +
\frac{1}{2}v_1^2\right) \equiv v\frac{dw}{dx} = - \mathcal{L}   \]
in the cooling gas, and $\gamma$ may be regarded as a variable to
take account of internal excitation.

The various processes giving rise to cooling behind shocks has
been treated in detail in two important papers by \citet{HM79} and
\citet{MH80}. Very roughly, in order of decreasing
temperature, the major coolants are dissociation, collisional
excitation of H Lyman levels, [OI] line emission, \hm, H$_2$O (to
which any free oxygen is rapidly converted in the hot gas) and CO
line emission.

Both processes, shocking and cooling, increase the entropy of the
gas (the first adiabatically) from which we infer the obvious: a
shock cannot run in the opposite sense. Furthermore, as we saw, a
shock travels supersonically with respect to the gas ahead.
Downstream the gas has to be subsonic with respect to the shock so
that the high pressure which drives it continues to reach the
front (the sound speed has been increased to a fraction of \vsh\
and the exit speed has been reduced to a fraction of that speed).

The quantity being differentiated in the last equation, $w$, is
the {\sl stagnation enthalpy\/}. If the net cooling function
$\mathcal{L}$ is expressible as a function of $\rho$ and $T$ then
the equation is integrable by straightforward quadrature.

\begin{figure}
\centering
\includegraphics*[height=75mm]{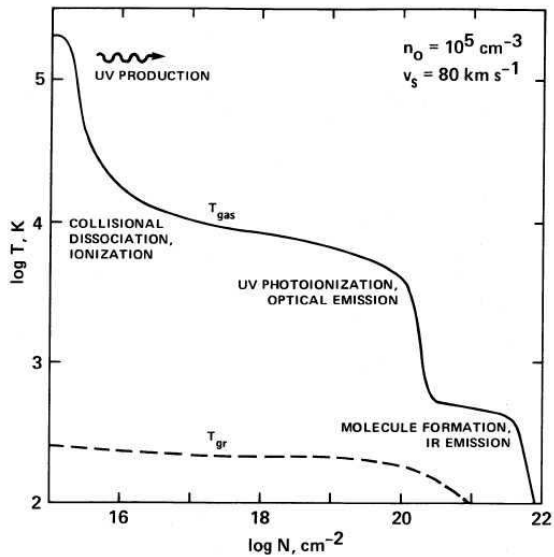}
\caption{(a) a fast ($80\;$\kms) fully dissociative J shock showing
$T$ as a function of column density $N$ through the shock. The
plateau at 10$^4\;$K is where the atomic gas is heated by UV
radiation from the front of the shock; the plateau at $400\;$K
extending to $N=10^{22}\;$cm$^{-2}$ is due to \hm\ reformation
\citep[from][]{HM89}.}
\label{fig:Jshock}
\end{figure}

If the shock is {\em strong\/}, {\em i.e.\/} the pressure ratio
$p_1/p_0$ or equivalently the initial Mach number $M =
\vsh$/(upstream sound speed) is large, then the R-H conditions
give (with $\gamma$ = 1.4, for diatomic molecules without other
internal states)
\[  \rho_1/\rho_0  =
    \vsh/v_1  =6,\hspace*{4mm} kT_1/\overline{m} =
    (5/36)\vsh^2,\hspace*{4mm} p_1 =(5/6)\rho_0\vsh^2.  \]

As a simple application of these ideas we can show how such shocks
have a characteristic velocity for the destruction of \hm. At the
highest temperatures the cooling is dominated by dissociation
until the temperature drops to a value $T_{\rm line}$ at which
line cooling dominates and dissociation is much reduced. If energy
$kT_{\rm diss}$ is released per dissociation at a rate $R$
kg$^{-1}$ s$^{-1}$ then the cooling rate is $\mathcal{L} = kT_{\rm
diss}R$. Each dissociation produces two H atoms with number
density $n(H)$, so $d(n(H)/\rho)/dt = 2R$, whence
\[   v\frac{d}{dx}\left(w + n(H)T_{\rm diss}/2\rho \right) = 0 \]
For a strong and initially molecular shock, we neglect the bulk
energy in the enthalpy, while the pressure is the sum of the
partial pressures of atomic hydrogen, molecular hydrogen and
helium (number fraction $y=0.1$), each a product of number density
and temperature. Define $f = n(H)/n_H$ where $n_H$ is the number
density of H nuclei and is proportional to $\rho$; use $\gamma =
5/3$ for atoms and $\gamma=7/5$ for molecules; and note that at
the start of the cooling zone $T=T_{\rm max}$ and $f=0$. Then it
is a matter of algebra to show that through the cooling flow
\[  f = \frac{(10y+7)(T_{\rm max}-T)}{2T_{\rm diss} + 3T}.   \]
If we take values of $52,000\;$K for $T_{\rm diss}$ and $T_{\rm line} =
3500\;$K, then $f$ becomes unity -- \hm\ is completely dissociated
-- at $T=T_{\rm line}$ when \vsh=24 \kms, the \citet{Kw77}
result.

In the case where we can treat the cooling zone as small compared
to other length scales in the region we can lump it together with
the shock transition. The jump conditions across this `isothermal
shock' (for simplicity it is assumed that initial and final
temperatures - and internal states - of the gas are the same:
hence the name) are
\[\rho_1/\rho_0=M_{\rm iso}^2, \hspace*{4mm}p_1 = \rho_0\vsh^2.\]
Cooling allows the density to reach very high values set by the
`isothermal Mach number' $M_{\rm iso} =
\vsh/\sqrt{kT_0/\overline{m}}$. In the cooling zone the pressure
has increased by only 17\%.

 If a magnetic field is flux-frozen to the gas ({\em i.e.\/} a fully
ionized plasma) and the field is directed at right angles to the
shock velocity, then pressure is the sum of that of plasma and
magnetic field. The magnetic pressure is $B^2/2\mu_0$. The
effective mass density of the magnetic field for non-relativistic
shocks is negligible, and the shock is assumed to be strong. The
momentum equation is then
\[ p + B^2/2\mu_0 + \rho v^2  =  B_0^2/2\mu_0 + \rho_0\vsh^2 \]
 while mass conservation becomes $\rho v=\rho_0\vsh$.
 Since flux freezing is assumed, $B\propto \rho \propto v^{-1}$.
 Denote the Alfv\'{e}n speed by $V_A$ ($V_A^2 =
 B^2/2\mu_0\rho$), and the Alfv\'{e}nic Mach number by $M_A =
 \vsh/V_{A,0}$.

 Writing $\overline{v}=v/\vsh$ and dividing through by $\rho_0\vsh^2$
 the momentum equation becomes
 \[  (p/\rho_0\vsh^2) + (2M_A^2\overline{v}^2)^{-1} + \overline{v} =
 (2M_A^2)^{-1} + 1 \]
and if downstream the gas is cool enough that magnetic pressure
dominates \citep[for more general conditions see][]{RD90} then the
downstream value of $v/\vsh$ is
\[  \overline{v} = (1 + \sqrt{1+8M_A^2})/4M_A^2.   \]
For magnetic pressure domination downstream ({\em i.e.\/} the
final Alfv\'{e}n speed greater than the sound speed, assumed equal
to the initial sound speed for simplicity), assuming as we have
that the Mach number $M$ is large,
\[   M_A < \sqrt{2}M^2.      \]
In such shocks the cooling zone is supported by the magnetic field
stress at a density not hugely different from the initial density
(thus slowing cooling processes in comparison with that in the
high density achieved in the absence of magnetic field).

The detailed structure of J-shocks is modified by the possible
changes of state of the material, and its effect on the
surroundings. If the shock is strong enough,  radiation generated
in this region will propagate ahead of the shock and may alter its
state depending on shock velocity. And since hydrogen molecule
re-formation is a slow process, dissociated molecules will not
reform in the close vicinity of the shock, and a separate
re-formation zone may occur \citep{HM89}.

\subsection{C Shocks}
\label{subsec:Cshocks}

 A quarter of a century ago, following an idea by \citet{Mu71},
\citet{Dr80} pointed out that, since molecular clouds are
slightly ionized and are permeated by magnetic fields, a different
kind of shock is possible and might be prevalent. This he named
the C shock, using the name J shock for the hydrodynamic
discontinuity just described.

The C shock can be envisaged in a two fluid continuum. One of the
fluids is the molecular neutral gas representing most of the mass,
while the other fluid is a conducting magnetized plasma consisting
of the tiny fraction (typically 10$^{-7}-10^{-5}$) of particles
ionized by cosmic rays plus the magnetic field inferred from
observation, in which the plasma is presumed to be flux-frozen. If
a smooth pressure gradient is generated perpendicular to the
magnetic field it will drive a compressive magnetohydrodynamic
wave through the plasma. This wave, if unencumbered by interaction
with neutral particles, would rapidly steepen into a shock
travelling faster than the magnetosonic speed (in these cases
roughly equal to the ionic Alfv\'{e}n speed, much greater than
either the thermal sound speed in the neutral gas or the
Alfv\'{e}n speed $M_A$ of the whole medium treated as a single
magnetized fluid). But by virtue of gas viscosity the magnetic
pressure is transferred slowly to acceleration of the neutral
material until far downstream the two components have the same
velocity. The viscous drag can ensure that the plasma pressure
wave will not accelerate to a shock state. For a wide range of
conditions (magnetic field strength, state of ionization) the
neutral gas is accelerated {\em slowly\/}, and remains supersonic
with respect to the structure throughout, without passing through
a hydrodynamic shock. The whole process reaches a stationary state
for suitable initial conditions far upstream and boundary
conditions downstream. The relatively wide region where this
transition occurs is the C shock. In this region the neutral gas
is heated by the viscous friction, and this region radiates.

\begin{figure}
\centering
\includegraphics[height=8cm]{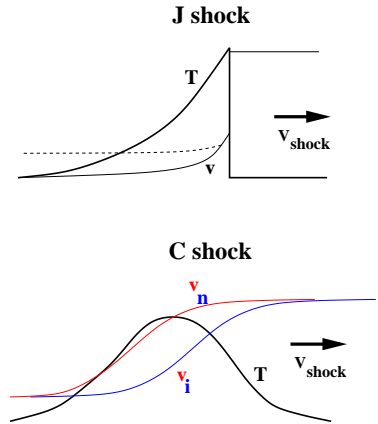}
\caption{Sketch, in the shock frame, of a J shock showing $T$ and
$v$, the dashed line corresponding to the effect of the presence
of a magnetic field, and a C shock, showing the slab of gas at a
relatively uniform temperature.} \label{fig:JandCshock}
\end{figure}

The structure of this type of shock is most easily considered
using the mass and momentum conservation equations introduced in
the discussion of J shocks, applied separately to the neutrals and
to the magnetized ion fluid.

This time the neutral fluid has an extra force on it caused by
friction with the ions, and vice versa. But in the summed momentum
equation these terms cancel, and we can use the equation
neglecting these forces. Generalizing the result for a magnetized
one-fluid shock above by noting that flux freezing applies to the
ions only, and denoting ion properties by subscript $i$ and
neutrals by $n$,
\[  (p_n/\rho_{n,0}\vsh^2) + (2M_A^2\overline{v}_i^2)^{-1} +
     \overline{v}_n = (2M_A^2)^{-1} + 1 \]
 whence we find a relationship between the ion and neutral
 velocities.

 If the cooling is such that thermal pressure is unimportant, then
 approximately
 \[   \overline{v}_n = 1 - (\overline{v}_i^{-2}-1)/2M_A^2.   \]
Far downstream where $v_i=v_n$ we have the same conditions as the
cold magnetized shock discussed above.

The slip speed, that is $v_i-v_n$, determines the amount of
frictional force and frictional heating in the flow. In the simple
case here we find the maximum value~\citep{Sm91b}
\[  (\overline{v}_i-\overline{v}_n)_{\rm max} = 1+1/2M_A^2 -
3/2M_A^{2/3}.  \]

\begin{figure}
\centering
\includegraphics[height=7cm]{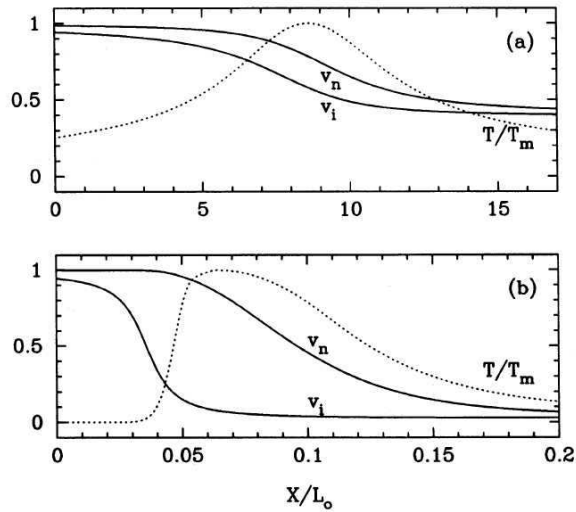}
\caption{(a) A low $M_A$ C shock showing $v_i$ and $v_n$ changing
in parallel. (b) A high $M_A$ C shock sowing the rapid change in
$v_i$ and the consequent large velocity difference \citep[from][]{Sm91a}.} 
\label{fig:Cshocks}
\end{figure}

Consider the length scales in this shock structure. The frictional
body force (per unit volume) imposed on the neutrals by the ions,
and {\em vice versa,\/} has magnitude $F =
\rho_n\rho_i\langle\sigma v\rangle(v_n-v_i)/(m_i+m_n)
$~\citep{DMcK93}, where $\langle\sigma v\rangle$ is the ion-neutral
elastic collision rate, and subscripts $n,i$ have the obvious
meaning. The equations of motion for ion fluid (neglecting ion
thermal and ram pressure) and neutrals (neglecting thermal
pressure) respectively are
\[  \frac{d}{dz}\left(\frac{B^2}{2\mu_0}\right) = F  \]
\[  \frac{d}{dz}\left(\rho v_n^2)\right) = -F.  \]
If we define $L_0 = \vsh / n_{i,0}\langle\sigma v\rangle$, roughly
the initial mean free path of neutrals through the ions times Mach
number, and define also the ion length $L_i = v_i/(dv_i/dz)$
and neutral length $L_n = v_n/(dv_n/dz)$, the first equation above
gives $L_i/L_0 \simeq M_A^{-2}$ and we also find from the
$(\overline{v}_i,\overline{v}_n)$ relation that $L_n / L_i = M_A$.

These relations for slip speed and length scales show the
following results. If $M_A$ is small the slip speed is a small
fraction of the shock velocity. Such soft shocks can accelerate
the neutrals without getting them very hot. The change in the ion
velocity closely parallels that in the neutrals. The length of the
slip zone is proportional to \vsh, and so can accelerate the
radiating molecules to quite large velocities.

 On the other hand a
high value of $M_A$ will create a flow in which the ions rapidly
approach their final speed before the neutrals have been
significantly decelerated, exposing the neutrals to a high rate of
high velocity collisions. There are two major consequences. The
heating rate is increased, and the gas may become sufficiently hot
that it becomes subsonic - which engenders a hydrodynamic jump
(the neglected neutral thermal pressure term in the equation above
becomes important). And in these conditions at a shock velocity of
40-50 \kms\ the neutrals can be ionized by single collisions,
leading to an ionization run-away as the increasing ion fraction
reduces $L_0$.

In any case if the length of the drag zone is too short compared
with the cooling length, the sound speed will increase and the gas
may become subsonic through a hydrodynamic jump. Thus, for a
limited range of conditions a C shock structure will contain a J
shock which may provide much of the cooling via high excitation
lines but still produce C shock-like emission of \hm\ for example.

The heated region behind each kind of shock is shown schematically
in Fig. \ref{fig:Cshocks}.

It is evident that, in contrast to J shocks, C shocks create a
zone that very roughly corresponds to a slab of gas at a given
temperature, and that therefore the emission will reflect that
status. That this is not shown in many of the observations is the
reason for proposing bow shocks as emitters \citep{Sc75,HR84,Mu87}.

\subsection{Bow Shock Structure}

There are few instances of observations of planar shocks. Usually
the next best guess is that one is observing a shock driven into a
bowed shape by a supersonic gas jet or supersonically moving dense
clump. The first models of Herbig Haro objects were of this
nature, and expose the essential complications.

The range of excitations and breadth of velocity profiles seen in
quite small projected areas on the sky led to the idea that one is
often observing some or indeed all of a bow shock structure. In
some special cases the bow is quite discernible.

A simple but perhaps suspect way in which to create models of such
regions is to use oblique planar models, and patchwork
these onto the surface of a pre-decided bow geometry. This
technique was pioneered by \citet{Ha87} for
optically emitting bows, and applied to \hm~\citep{Sm91a} to
explain the extreme conditions at the Peak 1 in the Orion outflow.

The strongest section of the shock is clearly at the head of the
bow. Here, the shock may be strong enough to ionize and dissociate
the molecules, and is certainly a J-shock. Further round the side
of the bow the shock may still be a J-shock but not completely
dissociating the molecules, while further round, the shock becomes
a C-shock~\citep{Sm90c}.

Clearly, even if the initial magnetic field had a simple geometry,
the varying angle of incidence at each part of the shock makes
such a structure hard to model. The extended length scales and
subtleties of structure of oblique C shocks \citep{WD87} make this
region the least straightforward part of the modelling process.

Notwithstanding, \citet{Sm91a} were able to
reproduce both the excitation (up to very high energies) and
the extreme breadth of the lines with a very low Alfv\'{e}n Mach
number and very fast shock. The lingering problem with this
explanation has always been seen to be the extremely high
pre-shock magnetic field required.

The technique is to specify an axially symmetric bow shape
$R=f(z)$ ($z$ is distance along the axis; $R$ is perpendicular
distance from axis of the bow at $z$) supposedly created by a
uniform supersonic flow, velocity \vsh, hitting a small obstacle,
or equivalently by such an obstacle -- or indeed the end of a jet
-- ramming supersonically into a uniform medium. If $dz/dR =
\tan\psi$ the effective shock speed at a point in the bow is $\vsh
\cos\psi$ and the shocked material has, in addition to its
post-shock behaviour perpendicular to the shock, a component of
velocity $\vsh \sin\psi$ tangential to the bow. A line of sight
and a projected area around it are assumed to model an
observation. At each point of the bow within that area a line
intensity (per unit area of shock) through the shock flow is
calculated, as are the projection factor (secant of angle between
line of sight and shock normal), and the velocity component along
line of sight. From this the line profile is found, the shape and
intensity of the line being diagnostic of the shock properties.

\section{Models and Observations}

\subsection{Issues}
\label{subsect:issues}

Much of the theoretical work until recently has been based on the
assumption that the flow is stationary, and that timescales for
chemical evolution were long, and timescales for population of
internal states were short, and that upstream conditions are
static. Even now the modelling of shocks, including chemistry, grain
physics and electrodynamics and the effects of the dust on the
chemistry and dynamics, and obliquity progresses inevitably
slowly. The fact that post-shock turbulence will occur in many of the
flow structures that have been measured has led to some pioneering
studies of the process, but much remains to be done.

The question of how to extend a two-fluid approximation to account
for the behaviour and effects of the dust grain population has
been addressed several times, and because of the still uncertain
properties of grains is still an open question.

Even in plane steady normal shocks several instabilities have been
discovered. Are there more? Is it possible to make useful
predictions from unstable shocks?

On the observational side, the ISO measurements, opening wider the
near to far IR spectral region, have produced a plethora of new
results. But there is still the usual urgent need of better
spatial resolution (and of course signal-to-noise ratio) to
compensate for our inability to see round the side of these
complex regions.

\subsubsection{Grain Charge and Drag}

The picture so far of the C-shock is of a two-fluid process, but
the presence of charged dust grains can significantly alter the
picture, particularly in cold high density regions~\citep{Dr80}.
Several papers investigate the tricky issue of grains as a
subspecies in the shock flow~\citep{Pi90,Ca97,Wa98,FP03}. Also see the
chapter by Tom Hartquist and Ove Havnes.

The first issue is that the grains may be charged, and can be the
dominant ionized species providing the drag on neutrals. This will
occur preferentially at high densities. \citet{Pi90} found that the
dust in high density shocks was constrained by charge separation to
move at a speed intermediate between that of the ions and of the
neutrals, with the net effect of steepening the shock
structure. \citet{Wa98} examined the possible flows in great detail,
and pointed out that the effect is accentuated by taking the
out-of-plane deviations of the magnetic field into account, thus
finding even steeper (and by implication hotter) shock
structures. These effects are greatest at high density and in the
presence of small grains (which for a given grain mass will give
greater drag). \citet{Wa98} noted that, since most C shocks will give
temperatures between 1000K and 2000K, and molecules at that
temperature will radiate most of the shock energy, there may be no
dominant observed effect in changing the detailed structure of the C
shock. Conversely, he pointed out that reliable diagnostics will rely
on getting the physical processes correct in detail.

\citet{FP03} addressed some of the physical and chemical detail. They
show that large grains rapidly charge by electron attachment in the
shock flow and become attached to the magnetic field, and that the
critical velocity (at which H dissociation leads to breakdown to a J
shock) is sufficiently high \citep{Le02} that refractory grains can
shatter. \citet{Ch06} examined the effects of obliquity of the
magnetic field, and emphasized the significance of the grain physics
in determining the flow structure. \citet{Ci04} pointed out that the
degree to which the grains attach to the magnetic field determines the
ionic magnetoionic sound speed, and if a significant fraction of the
total grain mass is thus attached, the ionic material will shock,
destroying the C shock.

\subsubsection{Grain Destruction}

In all of the calculations referred to above there is the issue of
grain destruction (or more positively element release) either by
sputtering or shattering of dust grains.

Calculations by \citet{Dr95} and \citet{FP95} have shown that C shocks
can destroy dust grain mantles and cores, and \citet{Ca97} have shown
that such shocks may release an abundance of Si consistent with
observations of shocked regions, and many orders of magnitude above
the average abundance. \citet{Ca97} emphasized that destruction is
more efficient in oblique shocks than in perpendicular shocks.  This
process may be the dominant producer of SiO seen in high density star
forming regions and indicated in the observations of \citet{Ji05}, for
example.

\citet{FP03} demonstrated that a
specific C shock travelling at 50 \kms\ would shatter over half of
the amorphous carbon grains present in the model.

All of these findings together imply that the behaviour of C
shocks is dependent on dealing with chemistry, dust grain physics
and MHD phenomena in great detail (since some of the necessary
grain physics remains to be determined, this is a worry).

\subsubsection{Chemistry and Time Dependence}

Another issue which has been clarified by the refinement of shock
models is that the processes determining the shocks have
timescales which are commensurate with shock passage times, but
also may be comparable with the time since the shock was created.
This raises the possibility for the pessimist that every shock
observed is unique (or rather that the number of determining
parameters, now including initial conditions and time since
creation, is far greater than observations can encompass). At any
rate it emphasizes how important the detail can be in determining
the overall effect.

Further, if the pressure driving the shock changes, or the medium
into which it propagates has large gradients, or is lumpy,
externally imposed time dependent effects come into play.

An example of what time-dependent driving can do is shown by \citet{Li02}. 
They demonstrated that a shocked layer
driven by a slowly accelerating piston can accumulate molecular
gas without destroying the \hm, producing column densities and
velocities of gas comparable with those observed in the Orion
outflow in an acceleration period of tens of years, and protecting
the molecular layer from the hot gas at the shock by an
intermediate layer of H atoms.

\citet{Gi05} have demonstrated that near HH2,
and caused by it, a wide variety of different excitation phenomena
can be created by a shock in a non-uniform medium.

\cite{Ch98} demonstrated that the
timescale for an initial disturbance in a dark cloud to reach
steady state is so long that it may exceed the lifetime of the
outflow causing the shock. They showed further that during the
evolution to a C shock there will be a J-shock embedded, and
that the combined radiation from this region and from the rest of
the (C type) flow may explain the \hm\ observations in some
sources where several different excitation temperatures are
measured. They emphasized again the necessity of calculating in
detail the degree of ionization in particular.

These more detailed studies also reveal that even if a J shock
occurs all \hm\ emission may not be lost! \citet{Fl03} and \citet{Ca04}) 
demonstrated that for fully
dissociating shocks of moderate speed an \hm\ emission spectrum
can be produced which can account for some of the observations.

The issue of time dependence is intimately tied up with
instabilities. The most marked(!) instability is the
\citet{Wa90} instability of plane C shocks, in which, rather
like the Parker instability, ripples on the lateral magnetic field
lines accumulate a higher density of ions and are dragged further
downstream, trapping higher density gas which cools rapidly. The
astonishing thing about this instability, which also occurs in
oblique C shocks~\citep{Wa91}, is that it has little effect on the
overall output from the shock, for reasons touched on before: the
hot gas does the radiating~\citep{MS97,NS97}.

Many of the developing models reveal one dimensional chaotic or
bouncing instabilities, due for example to shortened cooling times
in dissociation or ionization zones behind
shocks~\citep{Li02,SR03,Le04}. ``What will
happen in three dimensional models?'' is an open question.

\subsubsection{Turbulence and Turbulent Mixing and Other Effects}

\citet{CR91}, in a ground-breaking paper, investigated heating effects
of a turbulent mixing layer. \citet{TR95} showed that such zones could
mimic post-shock flows in some respects. Amongst others, \citet{Pa02},
\citet{ES04}, and \citet{He05} have begun to model more general cloud
turbulence as a source of structure and radiation. This work is still
at an exploratory stage, and in future may converge with that on shock
codes.

\subsection{Current Models and Observations}

In the interstellar medium in our Galaxy we expect that the
cooling zone in most observed shocks will be $\sim 10^{-5}$ pc in
the case of J-shocks and $\sim 10^{-2}$ pc for C-shocks in very
round numbers. This means that it is unlikely that J-shocks will
be resolved, while C-shocks can be resolved with current
equipment.

The arrival of large masses of ISO satellite data has enlivened
this area~\citep{vD04}.

Modelling is rapidly becoming more sophisticated, with attention being
paid to optimum methods \citep{Fa03,Le04}, and more of the required
physics (grains, chemistry, time dependence) is being included. What
has become clear \citep{Ch98,Fl03,Le04,Ca04}) is that rugged
predictions require very complete modelling.

Now a few examples are considered.

The extensive line list of \hm\ from HH43, observed by \citet{Gi02},
has been fit by \citet{Fl03} by either a 25 \kms\ J shock {\em or\/}
an 80 \kms\ C shock! The same modellers fit data by \citet{Wr96} and
\citet{Fr02} of Ceph A West with a model of a J shock with a C
precursor typical of early evolution.

\citet{OC05} modelled the \hm\ emission
structures seen in the HH211 outflow using C bow shocks.

\citet{Sn05} examined the supernova remnant
IC443 using the Submillimeter Wave Astronomy Satellite (SWAS) to observe
H$_2$O lines. They concluded that the data are consistent with a
combination of fast J shocks and slow (J or C) shocks, perhaps
likely in a clumpy medium.

The Orion outflow, pictured in part in Fig. \ref{fig:OrionShock},
has been a test-bed and conundrum throughout the entire
development of this area.

\citet{Le02} produced the \hm\ excitation diagram which matches the
observed data of \citet{Ro00} by superposing two C shocks with
velocities as high as 40 \kms\ and 60 \kms.

The discovery of the `bullets' or fingers~\citet{AB93} \citep[see
also][]{MM96} suggested an explosive event, and \citet{St95} provided
an explanation for the major structural features (in particular the
fingers) by having a time variable outflow, from IRc2 or the BN
object, which sweeps up a shell and then accelerates it whereupon it
fragments via Rayleigh-Taylor instability.

There has been a great deal of progress in establishing the values
for important rates in the microphysics and chemistry, in
developing models that can cope with the subtleties of grain
dynamics and time-dependent chemistry, and in the observations
(particularly thanks to ISO in the mid IR); but an enormous amount
remains to be done.

%
%

%
%

%
%
%
%
%

\printindex

\end{document}